\documentstyle[12pt]{article}

\begin{document}

\begin{center}
{\Large {\bf Induced scattering and two quantum absorption of Alfv\'{e}n
waves with arbitrary angles of propagation.}} \footnote{%
To be published in JETP {\bf 90}(5),810,(2000)} 

V.N.Zirakashvili

{\it Institute for Terrestrial Magnetism, Ionosphere and Radiowave
Propagation, 142190, Troitsk, Moscow Region, Russia}

{\bf Abstract}

{\small The equation for temporary evolution of spectral energy of
collisionless Alfv\'{e}n waves is derived in framework of weak turbulence
theory. The main nonlinear processes for such conditions are induced
scattering and two quantum absorption of Alfv\'{e}n waves by thermal ions.
The equation for velocity distribution of thermal particles is derived that
describes diffusion in momentum space due to this nonlinear processes.
Comparison is done with the results of another authors. Results obtained are
qualitatively differ from the ones obtained for the case of Alfv\'{e}n waves
propagation along mean magnetic field.}
\end{center}

{\bf Introduction}

It is well known that Alfv\'{e}n waves are very important agent of cosmic
plasma. It is really believed now that main part of energy of interplanetary
and interstellar magnetohydrodynamic (MHD) turbulence is contained in Alfv%
\'{e}n waves. This is because other types of low frequency waves - ionsound
and magnetosonic waves are strongly damped due to linear Landau damping. For
Alfv\'{e}n waves such a damping is relatively small and nonlinear effects
should be taken into account in order to investigate dissipation of MHD
turbulence. It is well known now that the main nonlinear processes for such
waves are induced scattering and two quantum absorption by thermal ions.
Corresponding damping rates were calculated by many authors [1-7].
Nevertheless totally comprehensive study is absent now. Firstly induced
scattering of Alfv\'{e}n waves was considered in paper [1] but for $\beta
\ll 1$. Here $\beta $ is the square of the ratio of thermal ion velocity $%
v_{Ti}=\sqrt{T_{i}/m_{i}}$ ($T_{i}$ - is ion temperature in energetic units, 
$m_{i}$ is ion mass) and Alfv\'{e}n velocity $v_{a}=B_{0}/\sqrt{4\pi \rho
_{0}}$ ($B_{0}$ is the mean magnetic field strength, $\rho _{0}$ is the mean
plasma density). In papers [2-5] the case of wave propagation along mean
magnetic field was considered. At last in papers [6,7] the oblique
propagation of Alfv\'{e}n waves was considered but random magnetic field
component along mean magnetic field arising from magnetic pressure of Alfv%
\'{e}n waves was not taking into account. As will be shown in this paper the
last effect strongly changes resultant damping rates. In particular, the
interaction of waves with the same signs of wavevector components along mean
magnetic field is impossible.

{\bf Basic equations.}

Kinetic approach is the most adequate method in nonlinear plasma theory. In
framework of weak turbulence it assumes the expansion of velocity
distribution of thermal particles in series on powers of random electric and
magnetic fields and calculation of nonlinear electric currents which are
substituted in Maxwell equations. Such a procedure applied for Alfv\'{e}n
waves results in cumbersome expressions containing multiple series of Bessel
functions (cf. [1]). If the case of magnetized thermal particles is
considered this functions should be expanded on powers of theirs small
argument. These calculations are very tedious. We shall use another approach
here that seems also more attractive from physical point of view. If
wavelengths are much larger then thermal particles gyroradii it is possible
to use equations averaged on gyrorotation. This means that we shall neglect
by all dispersive effects related with finite particle gyroradius. The
system of equations can be written as follows [8]: 
\begin{equation}
\rho \left( \frac{\partial {\bf u}}{\partial t}+({\bf u\nabla )u}\right) =-%
\frac{1}{4\pi }\left[ {\bf B\times }[\nabla \times {\bf B]}\right] -\nabla
P_{\perp }-(\nabla {\bf b}){\bf b}(P_{\parallel }-P_{\perp })
\end{equation}
\[
\frac{\partial F_{\alpha }}{\partial t}+({\bf u}_{E}+v_{\parallel }{\bf b}%
)\nabla F_{\alpha }-\frac{v_{\perp }}{2}\frac{\partial F_{\alpha }}{\partial
v_{\perp }}\left( v_{\parallel }\nabla {\bf b}+\nabla {\bf u}_{E}+{\bf u}%
_{E}({\bf b}\nabla ){\bf b}\right) +
\]
\begin{equation}
+\frac{\partial F_{\alpha }}{\partial v_{\parallel }}\left( \frac{q_{\alpha }%
}{m_{\alpha }}E_{\parallel }+\frac{v_{\perp }^{2}}{2}\nabla {\bf b}+{\bf u}%
_{E}\left( v_{\parallel }({\bf b}\nabla ){\bf b}+({\bf u}_{E}\nabla ){\bf b}+%
\frac{\partial {\bf b}}{\partial t}\right) \right) =0
\end{equation}
\begin{equation}
\frac{\partial {\bf B}}{\partial t}=[\nabla \times \lbrack {\bf u}\times 
{\bf B}]]
\end{equation}
\begin{equation}
\nabla {\bf B}=0
\end{equation}
\begin{equation}
\sum\limits_{\alpha }q_{\alpha }\int d^{3}vF_{\alpha }=0
\end{equation}
where ${\bf {b=B/}}B$ is a unit vector along magnetic field ${\bf {B}}$, $%
{\bf {u}}_{E}{\bf =}c{\bf [{E\times B}]/}B^{2}$ is electric drift velocity, $%
E_{\parallel }$ is electric field component along magnetic field, $\rho $
and ${\bf {u}}$ are plasma density and velocity, $F_{\alpha }(v_{\parallel
},v_{\perp })$, $q_{\alpha }$, $m_{\alpha }$ are velocity distribution,
electric charge and mass of thermal species $\alpha $. Perpendicular and
parallel pressures in Eq. (1) are given by the formulae 
\begin{equation}
P_{\parallel }=\sum\limits_{\alpha }\int d^{3}vm_{\alpha }(v_{\parallel }-%
{\bf ub})^{2}F_{\alpha },
\end{equation}
\begin{equation}
P_{\perp }=\sum\limits_{\alpha }\int d^{3}v\frac{m_{\alpha }v_{\perp }^{2}}{2%
}F_{\alpha }
\end{equation}

Guiding center equation (2) takes into account kinetic effects of thermal
particles. For the case of frozen magnetic field that is considered here
electric drift velocity ${\bf {u}}_{E}$ is simply perpendicular to magnetic
field component of medium velocity ${\bf {u}}$.\newline
\qquad {\bf Expansion on powers of medium velocity. }\newline
\qquad Let mean magnetic field ${\bf {B}_{0}}$ is in $z$-direction. Let us
write all quantities in a form ${\bf {B=B}_{0}+\delta {B}}$, $\rho =\rho
_{0}+\delta \rho $, etc. and expand them in Fourier series: $\delta {\bf B}%
=\sum\limits_{k}{\bf B}_{k}\exp (i{\bf kr}-i\omega t)$, $\delta \rho
=\sum\limits_{k}\delta \rho _{k}\exp (i{\bf kr}-i\omega t)$ etc., $k=({\bf {%
k,}}\omega {\bf )}$. All these Fourier components should be expanded on
powers of medium velocity due to presence of linear polarized Alfv\'{e}n
waves. As for linear theory of Alfv\'{e}n waves $\delta \rho _{k}=0$, $%
E_{\parallel k}=0$, $B_{zk}=0$, $u_{zk}=0$, these quantities should be
considered as second order ones. Performing \ expansion in Eq. (2) one can
obtain second order $\delta F_{\alpha k}$. Substitution of this quantity
into quasineutrality equation (6) permits to find $E_{\parallel k}$. As a
result 
\[
\delta F_{\alpha k}=\frac{B_{zk}}{B_{0}}\left[ \frac{k_{z}}{v_{\parallel
}k_{z}-\omega }\left( \frac{v_{\perp }^{2}}{2}-\frac{q_{\alpha }}{m_{\alpha }%
}\sigma _{1}(k)\right) \frac{\partial F_{0\alpha }}{\partial v_{\parallel }}-%
\frac{v_{\perp }}{2}\frac{\partial F_{0\alpha }}{\partial v_{\perp }}\right]
+
\]
\[
+\frac{1}{2}\sum\limits_{k=k^{\prime }+k^{\prime \prime }}({\bf u}%
_{k^{\prime }}{\bf u}_{k^{\prime \prime }})\left[ -\frac{k_{z}^{\prime
}k_{z}^{\prime \prime }}{2\omega ^{\prime }\omega ^{\prime \prime }}v_{\perp
}\frac{\partial F_{0\alpha }}{\partial v_{\perp }}+\right. 
\]
\begin{equation}
\left. +\frac{k_{z}}{v_{\parallel }k_{z}-\omega }\frac{\partial F_{0\alpha }%
}{\partial v_{\parallel }}\left( \frac{v_{\perp }^{2}}{2}\frac{k_{z}^{\prime
}k_{z}^{\prime \prime }}{\omega ^{\prime }\omega ^{\prime \prime }}+\frac{%
v_{\parallel }}{k_{z}}\left( \frac{k_{z}^{\prime 2}}{\omega ^{\prime }}+%
\frac{k_{z}^{\prime \prime 2}}{\omega ^{\prime \prime }}\right) -1+\frac{%
q_{\alpha }}{m_{\alpha }}\sigma _{0}(k^{\prime },k^{\prime \prime })\right) %
\right] ,
\end{equation}
\begin{equation}
E_{\parallel k}=ik_{z}\left[ \frac{B_{zk}}{B_{0}}\sigma
_{1}(k)-\sum\limits_{k=k^{\prime }+k^{\prime \prime }}\frac{1}{2}\sigma
_{0}(k^{\prime },k^{\prime \prime })({\bf u}_{k^{\prime }}{\bf u}_{k^{\prime
\prime }})\right] .
\end{equation}

For simplicity of notation we shall omit mean value of the product ${\bf u}%
_{k^{\prime }}{\bf u}_{k^{\prime \prime }}$ that should be extracted from
correspondent product. Quantities $\sigma _{0}(k^{\prime },k^{\prime \prime
})$ and $\sigma _{1}(k)$ in expressions (8) and (9) can be expressed in
terms of velocity distribution of thermal particles: 
\[
\sigma _{0}(k^{\prime },k^{\prime \prime })=\left[ \sum\limits_{\alpha }%
\frac{q_{\alpha }^{2}}{m_{\alpha }}\int \frac{d^{3}v}{v_{\parallel
}(k_{z}^{\prime }+k_{z}^{\prime \prime })-\omega ^{\prime }-\omega ^{\prime
\prime }}\frac{\partial F_{0\alpha }}{\partial v_{\parallel }}\right]
^{-1}\cdot 
\]
\begin{equation}
\cdot \sum\limits_{\alpha }q_{\alpha }\int \frac{d^{3}v}{v_{\parallel
}(k_{z}^{\prime }+k_{z}^{\prime \prime })-\omega ^{\prime }-\omega ^{\prime
\prime }}\frac{\partial F_{0\alpha }}{\partial v_{\parallel }}\left( 1-\frac{%
v_{\perp }^{2}}{2}\frac{k_{z}^{\prime }k_{z}^{\prime \prime }}{\omega
^{\prime }\omega ^{\prime \prime }}-\frac{v_{\parallel }}{k_{z}^{\prime
}+k_{z}^{\prime \prime }}\left( \frac{k_{z}^{\prime 2}}{\omega ^{\prime }}+%
\frac{k_{z}^{\prime \prime 2}}{\omega ^{\prime \prime }}\right) \right)
\end{equation}
\begin{equation}
\sigma _{1}(k)=\left[ \sum\limits_{\alpha }\frac{q_{\alpha }^{2}}{m_{\alpha }%
}\int \frac{d^{3}v}{v_{\parallel }k_{z}-\omega }\frac{\partial F_{0\alpha }}{%
\partial v_{\parallel }}\right] ^{-1}\sum\limits_{\alpha }q_{\alpha }\int 
\frac{d^{3}v}{v_{\parallel }k_{z}-\omega }\frac{v_{\perp }^{2}}{2}\frac{%
\partial F_{0\alpha }}{\partial v_{\parallel }}
\end{equation}

Let us calculate $z$-component of random magnetic field $B_{zk}$ now. In
order to do this one should expand Fourier transform of Eq. (1,3) to the
second order and multiply Eq. (1) on \ ${\bf {k}_{\perp }}$. The final
result can be written as follows 
\[
\frac{B_{zk}}{B_{0}}=\sum\limits_{k=k^{\prime }+k^{\prime \prime }}\frac{(%
{\bf u}_{k^{\prime \prime }}{\bf k}^{\prime })({\bf u}_{k^{\prime }}{\bf k}%
^{\prime \prime })}{v_{a}^{2}\Delta (k)}\left( \frac{k_{z}^{\prime
}k_{z}^{\prime \prime }c_{a}^{2}}{\omega ^{\prime }\omega ^{\prime \prime }}%
-1\right) -\frac{1}{2}\sum\limits_{k=k^{\prime }+k^{\prime \prime }}({\bf u}%
_{k^{\prime }}{\bf u}_{k^{\prime \prime }})\frac{k_{z}^{\prime
}k_{z}^{\prime \prime }}{\omega ^{\prime }\omega ^{\prime \prime }}-
\]
\[
-\sum\limits_{k=k^{\prime }+k^{\prime \prime }}\frac{({\bf u}_{k^{\prime }}%
{\bf u}_{k^{\prime \prime }})}{2v_{a}^{2}\Delta (k)}\left[ \frac{%
k_{z}^{\prime }k_{z}^{\prime \prime }}{\omega ^{\prime }\omega ^{\prime
\prime }}\left( \omega ^{2}-c_{a}^{2}k_{z}^{2}\right) +\right. 
\]
\begin{equation}
\left. +k_{\perp }^{2}\sum\limits_{\alpha }\frac{m_{\alpha }}{\rho _{0}}{%
\int \frac{d^{3}vk_{z}}{v_{\parallel }k_{z}-\omega }\frac{v_{\perp }^{2}}{2}%
\frac{\partial F_{0\alpha }}{\partial v_{\parallel }}\left( \frac{%
v_{\parallel }}{k_{z}}\left( \frac{k_{z}^{\prime 2}}{\omega ^{\prime }}+%
\frac{k_{z}^{\prime \prime 2}}{\omega ^{\prime \prime }}\right) -1+\frac{%
q_{\alpha }}{m_{\alpha }}\sigma _{2}(k^{\prime },k^{\prime \prime })\right) }%
\right] 
\end{equation}
Here quantity $\sigma _{2}(k^{\prime },k^{\prime \prime })$ is given by the
expression 
\[
\sigma _{2}(k^{\prime },k^{\prime \prime })=\left[ \sum\limits_{\alpha }%
\frac{q_{\alpha }^{2}}{m_{\alpha }}\int \frac{d^{3}v}{v_{\parallel
}(k_{z}^{\prime }+k_{z}^{\prime \prime })-\omega ^{\prime }-\omega ^{\prime
\prime }}\frac{\partial F_{0\alpha }}{\partial v_{\parallel }}\right]
^{-1}\cdot 
\]
\begin{equation}
\cdot \sum\limits_{\alpha }q_{\alpha }\int \frac{d^{3}v}{v_{\parallel
}(k_{z}^{\prime }+k_{z}^{\prime \prime })-\omega ^{\prime }-\omega ^{\prime
\prime }}\frac{\partial F_{0\alpha }}{\partial v_{\parallel }}\left( 1-\frac{%
v_{\parallel }}{k_{z}^{\prime }+k_{z}^{\prime \prime }}\left( \frac{%
k_{z}^{\prime 2}}{\omega ^{\prime }}+\frac{k_{z}^{\prime \prime 2}}{\omega
^{\prime \prime }}\right) \right) 
\end{equation}
and denominator 
\begin{equation}
\Delta (k)=k_{\perp }^{2}D_{0}(k)+(c_{a}^{2}k_{z}^{2}-\omega ^{2})/v_{a}^{2}
\end{equation}
corresponds to dispersion relation of magnetosonic wave $\Delta (k)=0$.%
\newline
The velocity $c_{a}=\sqrt{v_{a}^{2}-(P_{\parallel 0}-P_{\perp 0})/\rho _{0}}$
substitutes Alfv\'{e}n velocity $v_{a}$ in plasma with anisotropic pressure.
We consider here the case of firehose stable plasma when the expression
under square root is positive. The quantity $D_{0}(k)$ is given by the
expression 
\begin{equation}
D_{0}(k)=1+\frac{8\pi P_{\perp 0}}{B_{0}^{2}}+\frac{4\pi }{B_{0}^{2}}%
\sum\limits_{\alpha }m_{\alpha }\int d^{3}v\frac{k_{z}}{v_{\parallel
}k_{z}-\omega }\left( \frac{v_{\perp }^{2}}{2}-\frac{q_{\alpha }}{m_{\alpha }%
}\sigma _{1}(k)\right) \frac{v_{\perp }^{2}}{2}\frac{\partial F_{0\alpha }}{%
\partial v_{\parallel }}
\end{equation}
\newline

{\bf Calculation of damping rates.}\newline
\qquad Standard method of calculation of damping rates uses the expansion of
equations (1-3) up to the third order of medium velocity. We use here more
simple method that gives the same result. We shall derive equations for
thermal particles and find probability of scattering. Practically one should
derive quasilinear equation from Eq. (2). The only difference is that
derivatives on velocity in Eq. (2) are multiplied on factors of second or
higher order. This equation can be written as follows: 
\[
\frac{\partial F_{0\alpha }}{\partial t}=\frac{\partial }{\partial
v_{\parallel }}\sum\limits_{k}\left| \frac{B_{zk}}{B_{0}}\left( \frac{%
v_{\perp }^{2}}{2}-\frac{q_{\alpha }}{m_{\alpha }}\sigma _{1}(k)\right)
+\right. 
\]
\begin{equation}
\left. +\frac{1}{2}\sum\limits_{k=k^{\prime }+k^{\prime \prime }}({\bf u}%
_{k^{\prime }}{\bf u}_{k^{\prime \prime }})\left( \frac{v_{\perp }^{2}}{2}%
\frac{k_{z}^{\prime }k_{z}^{\prime \prime }}{\omega ^{\prime }\omega
^{\prime \prime }}+\frac{v_{\parallel }}{k_{z}}\left( \frac{k_{z}^{\prime
}{}^{2}}{\omega ^{\prime }}+\frac{k_{z}^{\prime \prime }{}^{2}}{\omega
^{\prime \prime }}\right) -1+\frac{q_{\alpha }}{m_{\alpha }}\sigma
_{0}(k^{\prime },k^{\prime \prime })\right) \right| ^{2}\frac{k_{z}^{2}}{%
i(v_{\parallel }k_{z}-\omega )}\frac{\partial F_{0\alpha }}{\partial
v_{\parallel }}
\end{equation}
Substituting expression (12) for $B_{zk}$ and performing assemble averaging
one can obtain 
\[
\frac{\partial F_{0\alpha }}{\partial t}=\frac{\partial }{\partial
v_{\parallel }}\sum\limits_{{\bf k}^{\prime },{\bf k}^{\prime \prime }}W(%
{\bf k}^{\prime })W({\bf k}^{\prime \prime })\left[ (k_{z}^{\prime
}+k_{z}^{\prime \prime })^{2}w({\bf k^{\prime }},{\bf k^{\prime \prime }}%
,\omega ({\bf {k}^{\prime }),\omega (k^{\prime \prime }))+}\right. 
\]
\begin{equation}
\left. +(k_{z}^{\prime }-k_{z}^{\prime \prime })^{2}w({\bf k}^{\prime },-%
{\bf k}^{\prime \prime },\omega ({\bf k}^{\prime }),-\omega ({\bf k}^{\prime
\prime }))\right] \frac{\partial F_{0\alpha }}{\partial v_{\parallel }}.
\end{equation}
Here we use the expression 
\begin{equation}
<{\bf u}_{k}{\bf u}_{k^{\prime }}>=\frac{\delta (k+k^{\prime })}{2\rho _{0}}%
\left( W({\bf k})\delta (\omega -\omega ({\bf k}))+W(-{\bf k})\delta (\omega
+\omega ({\bf k})\right) ,
\end{equation}
where $W({\bf {k)}}$ is spectral energy density of Alfv\'{e}n waves with
dispersion relation $\omega ({\bf {k)}=}c_{a}{\bf |k}_{z}{\bf |}$. It's
normalized such that 
\[
\frac{\left\langle \delta B^{2}\right\rangle }{4\pi }\frac{c_{a}^{2}}{%
v_{a}^{2}}=\sum\limits_{{\bf k}}W({\bf k}).
\]
First term in square brackets of expression (17) corresponds to two quantum
absorption, the second one corresponds to induced scattering. The
probability of two quantum absorption $w({\bf {k}}^{\prime },{\bf {k}}%
^{\prime \prime },\omega ({\bf {k}}^{\prime }{\bf {)},}\omega {\bf ({k}}%
^{\prime \prime }{\bf {)})}$ is given by the formula 
\[
w({\bf k}^{\prime },{\bf k}^{\prime \prime },\omega ({\bf k}^{\prime
}),\omega ({\bf k}^{\prime \prime }))=\frac{\pi }{4\rho _{0}^{2}v_{a}^{4}}%
\delta (\omega ({\bf k}^{\prime })+\omega ({\bf k}^{\prime \prime
})-v_{\parallel }(k_{z}^{\prime }+k_{z}^{\prime \prime }))\cdot 
\]
\[
\cdot \left| \frac{1}{\Delta (k^{\prime }+k^{\prime \prime })}\left( \frac{%
v_{\perp }^{2}}{2}-\frac{q_{\alpha }}{m_{\alpha }}\sigma _{1}(k)\right) %
\left[ 2k_{\perp }^{\prime }k_{\perp }^{\prime \prime }\sin ^{2}\varphi (%
{\bf k}_{\perp }^{\prime },{\bf k}_{\perp }^{\prime \prime })\left( c_{a}^{2}%
\frac{k_{z}^{\prime }k_{z}^{\prime \prime }}{\omega ^{\prime }\omega
^{\prime \prime }}-1\right) \right. -\right. 
\]
\[
-\cos \varphi ({\bf k}_{\perp }^{\prime },{\bf k}_{\perp }^{\prime \prime
})\left\{ \frac{k_{z}^{\prime }k_{z}^{\prime \prime }}{\omega ^{\prime
}\omega ^{\prime \prime }}\left( (\omega ^{\prime }+\omega ^{\prime \prime
})^{2}-c_{a}^{2}(k_{z}^{\prime }+k_{z}^{\prime \prime })^{2}\right) +\left( 
{\bf k}_{\perp }^{\prime }+{\bf k}_{\perp }^{\prime \prime }\right)
^{2}\cdot \right. 
\]
\[
\left. \left. \cdot \sum\limits_{\alpha }\frac{m_{\alpha }}{\rho _{0}}\int 
\frac{d^{3}v\left( k_{z}^{\prime }+k_{z}^{\prime \prime }\right) }{%
v_{\parallel }(k_{z}^{\prime }+k_{z}^{\prime \prime })-\omega ^{\prime
}-\omega ^{\prime \prime }}\frac{v_{\perp }^{2}}{2}\frac{\partial F_{0\alpha
}}{\partial v_{\parallel }}\left( \frac{v_{\parallel }}{k_{z}^{\prime
}+k_{z}^{\prime \prime }}\left( \frac{k_{z}^{\prime 2}}{\omega ^{\prime }}+%
\frac{k_{z}^{\prime \prime 2}}{\omega ^{\prime \prime }}\right) -1+\frac{%
q_{\alpha }}{m_{\alpha }}\sigma _{2}(k^{\prime },k^{\prime \prime })\right)
\right\} \right] +
\]
\begin{equation}
\left. +v_{a}^{2}\cos \varphi ({\bf k}_{\perp }^{\prime },{\bf k}_{\perp
}^{\prime \prime })\left( \frac{v_{\parallel }}{k_{z}^{\prime
}+k_{z}^{\prime \prime }}\left( \frac{k_{z}^{\prime 2}}{\omega ^{\prime }}+%
\frac{k_{z}^{\prime \prime 2}}{\omega ^{\prime \prime }}\right) -1+\frac{%
q_{\alpha }}{m_{\alpha }}\sigma _{2}(k^{\prime },k^{\prime \prime })\right)
\right| _{_{\omega ^{\prime \prime }=\omega ({\bf k}^{\prime \prime
})}^{\omega ^{\prime }=\omega ({\bf k}^{\prime })}}^{2}
\end{equation}
Here $\varphi ({\bf {k}_{\perp }^{\prime },{k}_{\perp }^{\prime \prime }{)}}$
is the angle between ${\bf {k}_{\perp }^{\prime }}$ and ${\bf {k}_{\perp
}^{\prime \prime }}$. The spectral energy density of Alfv\'{e}n waves
evolves according to the equation 
\begin{equation}
\frac{\partial W({\bf k})}{\partial t}=-2\Gamma ({\bf k})W({\bf k})
\end{equation}
where damping rate can be expressed in terms of velocity distribution of
thermal particles and probabilities of scattering and two quantum absorption
[9,10]: 
\[
\Gamma ({\bf k})=-\omega ({\bf k})\sum\limits_{\alpha ,{\bf k}^{\prime
}}m_{\alpha }W({\bf k}^{\prime })\int d^{3}v\left[ (k_{z}+k_{z}^{\prime })w(%
{\bf k},{\bf k}^{\prime },\omega ({\bf k}),\omega ({\bf k}^{\prime
}))+\right. 
\]
\begin{equation}
\left. +(k_{z}-k_{z}^{\prime })w({\bf k},-{\bf k}^{\prime },\omega ({\bf k}%
),-\omega ({\bf k}^{\prime }))\right] \frac{\partial F_{0\alpha }}{\partial
v_{\parallel }}
\end{equation}
It is easy to see from expression (19) that waves with the same signs of $%
k_{z}$ can not interact. It is not so if one tend to zero $k_{\perp
}^{\prime }$ and $k_{\perp }^{\prime \prime }$. This means that
one-dimensional case is particular one (see Discussion).\newline
It is possible to transform damping rate (21) containing probability (19) to
the more convenient for applications form: 
\[
\Gamma ({\bf k})=\frac{1}{4}\omega ({\bf k})Im\sum_{{\bf k}^{\prime },\pm }%
\frac{W({\bf k}^{\prime })}{B_{0}^{2}/4\pi }\left[ \left( 1-\frac{1}{%
c_{a}^{2}}\left( \frac{\omega \pm \omega ^{\prime }}{k_{z}\pm k_{z}^{\prime }%
}\right) ^{2}\right) ^{2}D_{1}(k\pm k^{\prime })\cos ^{2}\varphi ({\bf k}%
_{\perp },{\bf k}_{\perp }^{\prime })+\right. 
\]
\[
+\frac{({\bf k}_{\perp }\pm {\bf {k}_{\perp }^{\prime })^{-2}}}{\Delta (k\pm
k^{\prime })}\left\{ 2k_{\perp }k_{\perp }^{\prime }\left( 1-c_{a}^{2}\frac{%
k_{z}k_{z}^{\prime }}{\omega \omega ^{\prime }}\right) \sin ^{2}\varphi (%
{\bf k}_{\perp },{\bf k}_{\perp }^{\prime })\pm \right. 
\]
\begin{equation}
\left. \left. \pm \cos \varphi ({\bf k}_{\perp },{\bf k}_{\perp }^{\prime
})\left( 1-\frac{1}{c_{a}^{2}}\left( \frac{\omega \pm \omega ^{\prime }}{%
k_{z}\pm k_{z}^{\prime }}\right) ^{2}\right) \left( D_{2}(k\pm k^{\prime })(%
{\bf k}_{\perp }\pm {\bf k}_{\perp }^{\prime })^{2}-\frac{%
c_{a}^{2}k_{z}k_{z}^{\prime }}{\omega \omega ^{\prime }}(k_{z}\pm
k_{z}^{\prime })^{2}\right) \right\} ^{2}\right] _{_{\omega ^{\prime
}=\omega ({\bf k}^{\prime })}^{\omega =\omega ({\bf k})}}
\end{equation}
and quantities $D_{1}(k)$ and $D_{2}(k)$ can be expressed in terms of
velocity distribution of thermal particles: 
\begin{equation}
D_{1}(k)=-v_{a}^{2}\sum_{\alpha }{\frac{m_{\alpha }}{\rho _{0}}\int \frac{%
d^{3}vk_{z}}{v_{\parallel }k_{z}-\omega }\left( 1-\frac{q_{\alpha }}{%
m_{\alpha }}\sigma _{3}(k)\right) \frac{\partial F_{0\alpha }}{\partial
v_{\parallel }}}
\end{equation}
\begin{equation}
D_{2}(k)=-\sum_{\alpha }{\frac{m_{\alpha }}{\rho _{0}}\int \frac{d^{3}vk_{z}%
}{v_{\parallel }k_{z}-\omega }\left( 1-\frac{q_{\alpha }}{m_{\alpha }}\sigma
_{3}(k)\right) \frac{v_{\perp }^{2}}{2}\frac{\partial F_{0\alpha }}{\partial
v_{\parallel }}},
\end{equation}
where $\sigma _{3}(k)$ is given by the expression 
\begin{equation}
\sigma _{3}(k)=\left[ \sum\limits_{\alpha }\frac{q_{\alpha }^{2}}{m_{\alpha }%
}\int \frac{d^{3}v}{v_{\parallel }k_{z}-\omega }\frac{\partial F_{0\alpha }}{%
\partial v_{\parallel }}\right] ^{-1}\sum\limits_{\alpha }q_{\alpha }\int 
\frac{d^{3}v}{v_{\parallel }k_{z}-\omega }\frac{\partial F_{0\alpha }}{%
\partial v_{\parallel }}
\end{equation}
It should be noted that damping rate (22) can be derived directly from
expansion of Fourier transforms of Eq. (1-3) up to the third order of medium
velocity but takes more tedious algebra.\newline

{\bf Damping rates for Maxwellian plasma.}\newline
Strictly speaking collision term should be added to the right-hand side of
equation (17). It tends to make the velocity distribution Maxwellian. It
will be so if frequency of collisions is high enough. On the other hand
frequency of collisions can be small enough for validity of collisionless
approximation. For Maxwellian plasma nonlinear damping rate (22) can be
expressed in terms of function $J_{+}(x)$ [1]: 
\begin{equation}
J_{+}(x)=-i\sqrt{\frac{\pi }{2}}x\exp \left( -\frac{x^{2}}{2}\right) +\frac{x%
}{\sqrt{2\pi }}\int\limits_{-\infty }^{+\infty }\frac{dx^{\prime }}{%
x-x^{\prime }}\exp \left( -\frac{x^{\prime 2}}{2}\right) 
\end{equation}
It contains principal value integral.\newline
For the sake of simplicity we shall consider pure hydrogen plasma. Damping
rate is mainly determined by thermal ions. Input of thermal electrons is $%
\sqrt{m_{i}/m_{e}}$ times smaller. The dependence of quantities $D_{0}(k)$,$%
D_{1}(k)$, $D_{2}(k)$ on frequency and wavenumber is then reduced to
dependence on $x=\frac{\omega }{v_{Ti}\left| k_{z}\right| }$: 
\begin{equation}
D_{0}(x)=1+2\beta J_{+}(x)+\beta \frac{J_{+}^{2}(x)}{1+\frac{T_{i}}{T_{e}}%
-J_{+}(x)}
\end{equation}
\begin{equation}
D_{1}(x)=\beta ^{-1}\frac{T_{i}}{T_{e}}\frac{1-J_{+}(x)}{1+\frac{T_{i}}{T_{e}%
}-J_{+}(x)}
\end{equation}
\begin{equation}
D_{2}(x)=\left( 1+\frac{T_{i}}{T_{e}}\right) \frac{1-J_{+}(x)}{1+\frac{T_{i}%
}{T_{e}}-J_{+}(x)}
\end{equation}
It was assumed here that ions and electrons can be at different temperatures 
$T_{i}$ and $T_{e}$. We shall consider two extreme cases below.\newline
$\beta \ll 1$.\newline
For this case damping rate can be written in the form 
\[
\Gamma (k_{z},{\bf k}_{\perp })=\frac{\omega ({\bf k})}{4}\sum_{{\bf k}%
^{\prime }}\theta (-k_{z}k_{z}^{\prime })\frac{W({\bf k}^{\prime })}{%
B_{0}^{2}/4\pi }\left[ \frac{v_{a}^{2}}{v_{Ti}^{2}}\frac{T_{i}^{2}}{T_{e}^{2}%
}\frac{\cos ^{2}\varphi ({\bf k}_{\perp },{\bf k}_{\perp }^{\prime })\sqrt{%
\frac{\pi }{2}}x\exp (-x^{2}/2)}{\left( 1-ReJ_{+}(x)+T_{i}/T_{e}\right) ^{2}+%
\frac{\pi }{2}x^{2}\exp (-x^{2})}\right| _{x=\frac{\omega ({\bf k)-}\omega (%
{\bf k}^{\prime })}{v_{Ti}\left| k_{z}-k_{z}^{^{\prime }}\right| }}+
\]
\begin{equation}
\left. +\delta (4k_{z}k_{z}^{\prime }+({\bf k}_{\perp }+{\bf k}_{\perp
}^{\prime })^{2})\frac{\pi }{({\bf k}_{\perp }+{\bf k}_{\perp }^{\prime
})^{2}}\left( 4k_{\perp }k_{\perp }^{\prime }\sin ^{2}\varphi ({\bf k}%
_{\perp },{\bf k}_{\perp }^{\prime })-({\bf k}_{\perp }+{\bf k}_{\perp
}^{\prime })^{2}\cos \varphi ({\bf k}_{\perp },{\bf k}_{\perp }^{\prime
})\right) ^{2}\right] ,
\end{equation}
where $\theta (x)$ is step function.\newline
The first term in expression (30) describes induced scattering of Alfv\'{e}n
waves which can be interpreted as generation of ionsound wave by pair of Alfv%
\'{e}n waves that is absorbed by thermal particles. For the case considered
scattering is differential, that is interaction of Alfv\'{e}n waves with
close absolute values of $k_{z}$ is possible only. The second term describes
two quantum absorption of Alfv\'{e}n waves which can be interpreted as
generation of magnetosonic wave that is absorbed by thermal particles also.
Expansion of spectral energy density of Alfv\'{e}n waves in the vicinity of $%
-k_{z}$ and change of summation to integration results in 
\[
\Gamma (k_{z},{\bf k}_{\perp })=\frac{v_{a}k_{z}^{2}}{B_{0}^{2}/4\pi }\int
d^{2}k_{\perp }^{\prime }\left[ -\alpha _{1}\left( W(-k_{z},{\bf k}_{\perp
}^{\prime })+k_{z}\frac{\partial }{\partial k_{z}}W(-k_{z},{\bf k}_{\perp
}^{\prime })\right) \cos ^{2}\varphi ({\bf k}_{\perp },{\bf k}_{\perp
}^{\prime })+\right. 
\]
\begin{equation}
\left. +W\left( -\frac{({\bf k}_{\perp }+{\bf k}_{\perp }^{\prime })^{2}}{%
4k_{z}},{\bf k}_{\perp }^{\prime }\right) \frac{\pi k_{z}^{-2}}{16({\bf k}%
_{\perp }+{\bf k}_{\perp }^{\prime })^{2}}\left( 4k_{\perp }k_{\perp
}^{\prime }\sin ^{2}\varphi ({\bf k}_{\perp },{\bf k}_{\perp }^{\prime })-(%
{\bf k}_{\perp }+{\bf k}_{\perp }^{\prime })^{2}\cos \varphi ({\bf k}_{\perp
},{\bf k}_{\perp }^{\prime })\right) ^{2}\right] ,
\end{equation}
and quantity $\alpha _{1}$ is given by integral 
\begin{equation}
\alpha _{1}=\sqrt{\frac{\pi }{2}}\frac{T_{i}^{2}}{T_{e}^{2}}%
\int\limits_{-\infty }^{+\infty }dx\frac{x^{2}\exp (-x^{2}/2)}{\left(
1-ReJ_{+}(x)+T_{i}/T_{e}\right) ^{2}+\frac{\pi }{2}x^{2}\exp (-x^{2})}=\pi 
\end{equation}
First term in expression (31) corresponds to result obtained by Livshits and
Tsytovich [1]. The only difference is the value of factor $\alpha _{1}$
which depends on $T_{i}/T_{e}$ in [1]. It is because in [1] simplifying
assumption $x=0$ in the denominator of integral (32) was used. The result
obtained here is more correct and can be derived in framework of standard
magnetohydrodynamics. \newline
$\beta \gg 1$.\newline
For this case expression (22) can be reduced to 
\[
\Gamma (k_{z},{\bf k}_{\perp })=\frac{\omega ({\bf k})}{4}\frac{v_{Ti}^{2}}{%
v_{a}^{2}}\sum_{{\bf k}^{\prime }}\theta (-k_{z}k_{z}^{\prime })\frac{W({\bf %
k}^{\prime })}{B_{0}^{2}/4\pi }\cos ^{2}\varphi ({\bf k}_{\perp },{\bf k}%
_{\perp }^{\prime })\cdot 
\]
\begin{equation}
\cdot Im\frac{1-J_{+}(x)}{1+\frac{T_{i}}{T_{e}}-J_{+}(x)}x^{4}\left[ \frac{%
T_{i}}{T_{e}}+\frac{\left( 1+\frac{T_{i}}{T_{e}}\right) ^{2}\left(
1-J_{+}(x)\right) }{J_{+}(x)\left( 2+2\frac{T_{i}}{T_{e}}-J_{+}(x)\right) }%
\right] _{x=\frac{\omega ({\bf k)+}\omega ({\bf k}^{\prime })}{v_{Ti}\left|
k_{z}+k_{z}^{\prime }\right| }}
\end{equation}
The damping rate (33) is determined by two quantum absorption that is again
differential on absolute value of $k_{z}$. Performing expansion similarly to
previous case one can obtain 
\begin{equation}
\Gamma (k_{z},{\bf k}_{\perp })=\alpha _{2}\frac{v_{Ti}k_{z}^{2}}{%
B_{0}^{2}/4\pi }\int d^{2}k_{\perp }^{\prime }W(-k_{z},{\bf k}_{\perp
}^{\prime })\cos ^{2}\varphi ({\bf k}_{\perp },{\bf k}_{\perp }^{\prime })
\end{equation}
where $\alpha _{2}$ is given by the expression: 
\begin{equation}
\alpha _{2}=\int\limits_{0}^{+\infty }x^{2}dxIm\frac{1-J_{+}(x)}{1+\frac{%
T_{i}}{T_{e}}-J_{+}(x)}\left[ \frac{T_{i}}{T_{e}}+\frac{\left( 1+\frac{T_{i}%
}{T_{e}}\right) ^{2}\left( 1-J_{+}(x)\right) }{J_{+}(x)\left( 2+2\frac{T_{i}%
}{T_{e}}-J_{+}(x)\right) }\right] 
\end{equation}
Numeric integration shows that this quantity practically does not depend on $%
T_{i}/T_{e}$. Numeric value $\alpha _{2}(T_{i}=T_{e})=2.25$.\newline
{\bf Discussion.}\newline
The main result of this paper is the expression for nonlinear damping rate
(22). Results obtained differ from results [6,7] where oblique Alfv\'{e}n
waves were also considered. We take into account components of random field,
corresponding to magnetosonic waves. As magnetosonic waves are strongly
damped (especially in $\beta >1$ plasma, exceptions are quasiparallel and
quasiperpendicular propagation), these components should be considered as
second order quantities and expressed in terms of medium velocity in Alfv%
\'{e}n waves. Simply speaking we treat these quantities similarly to
parallel electric field component $E_{\parallel k}$. This approach is
correct if energy density of Alfv\'{e}n waves is not concentrated in
wavevectors at small angles: 
\begin{equation}
<\delta B^{2}>/B_{0}^{2}\ll k_{\perp }^{2}/k_{z}^{2}
\end{equation}
For the opposite relation one-dimensional results [2-5] and results [6,7]
are valid. The results of this papers can be readily obtained if one let $%
B_{zk}=0$ in Eq. (16) or tend $k_{\perp }$, $k_{\perp }^{\prime }$ to zero
in expressions (19) and (22): 
\[
\Gamma ({\bf k})=\frac{1}{4}\omega ({\bf k})Im\sum_{{\bf k}^{\prime },\pm }%
\frac{W({\bf k}^{\prime })}{B_{0}^{2}/4\pi }\cos ^{2}\varphi ({\bf k}_{\perp
},{\bf k}_{\perp }^{\prime })\left[ \left( 1-\frac{1}{c_{a}^{2}}\left( \frac{%
\omega \pm \omega ^{\prime }}{k_{z}\pm k_{z}^{\prime }}\right) ^{2}\right)
^{2}D_{1}(k\pm k^{\prime })-\right. 
\]
\begin{equation}
\left. -\frac{v_{a}^{4}k_{z}k_{z}^{\prime }{}^{2}}{\omega ^{2}\omega
^{\prime }{}^{2}}D_{0}(k\pm k^{\prime })-2\frac{v_{a}^{2}k_{z}k_{z}^{\prime }%
}{\omega \omega ^{\prime }}\left( 1-\frac{1}{c_{a}^{2}}\left( \frac{\omega
\pm \omega ^{\prime }}{k_{z}\pm k_{z}^{\prime }}\right) ^{2}\right)
D_{2}(k\pm k^{\prime })\right] _{_{\omega ^{\prime }=\omega ({\bf k}^{\prime
})}^{\omega =\omega ({\bf k})}}
\end{equation}
In particular interaction of waves with the same signs of $k_{z}$ is
possible in this case. It is described by the second term of damping rate
(37). Derivation of damping rate (37) shows that in order to obtain validity
condition for approach used in this paper one should compare terms  $%
(k_{z}\pm k_{z}^{\prime })^{2}-(\omega \pm \omega ^{\prime })^{2}/c_{a}^{2}$
and $({\bf k}\pm {\bf k}^{\prime })^{2}$, in denominator $\Delta (k\pm
k^{\prime })$ of damping rate (22). For the waves with the same singes of $%
k_{z}$ the fist term is equal to zero if dispersion relation for Alfv\'{e}n
waves is used. It does not equal to zero if one takes into account nonlinear
shift of frequency that is of the same order as nonlinear damping rate. This
procedure results in condition (36). \newline
For $\beta \ll 1$ linear damping of magnetosonic waves is smaller than for
the case $\beta >1$ and is due to Landau damping on thermal electrons [11]: 
\begin{equation}
\gamma =\sqrt{\frac{\pi }{8}}v_{Te}\frac{k_{\perp }^{2}}{|k_{z}|}\frac{m_{e}%
}{m_{i}}
\end{equation}
and approach used in this paper is valid for stronger condition: 
\begin{equation}
\frac{<\delta B^{2}>}{B_{0}^{2}}\ll \sqrt{\frac{m_{e}T_{e}}{m_{i}T_{i}}\beta 
}\frac{k_{\perp }^{2}}{k_{z}^{2}}
\end{equation}
For the opposite relation linear damping of magnetosonic waves can be
disregarded and results [1] are valid. Three wave interactions of Alfv\'{e}n
and magnetosonic waves [11] should be also taken into account in this case.%
\newline
{\bf References}\newline
1. Livshits,M.A., Tsytovich,V.N. 1970, Nuclear Fusion {\bf 10}, 241.\newline
2. Lee,M.A., V\"{o}lk,H.J. 1973, Astroph. Sp. Sci. {\bf 24}, 31.\newline
3. Achterberg A. 1981, Astron. Astroph. {\bf 98}, 161.\newline
4. Kulsrud, R.M. 1982, Phys. Scripta, {\bf 2}/1 177.\newline
5. Miller, J.A. 1991, Ap. J. {\bf 376}, 342.\newline
6. Achterberg,A., Blandford,R.D. 1986, MNRAS {\bf 218}, 551.\newline
7. Fedorenko,V.N., Ostryakov,V.M., Polyudov, A.N., Shapiro, V.D. Preprint N
1267 A.F.Ioffe Phys.Tech. Inst. Leningrad, 1988.\newline
8. Kulsrud,R.M. 1983, MHD Description of Plasma, in {\it Basic Plasma Physics%
}, eds. Galeev,A.A, Sudan,R.N., North-Holland Co., Amsterdam. \newline
9. Tsytovich, V.N. 1977. Theory of turbulent plasma, Consultants Bureau, New
York, London.\newline
10. Tsytovich,V.N., Shvartsburg,A.B., 1966, JETP {\bf 22},554.\newline
11. Akhieser,A.I., Akhieser,I.A., Polovin,R.V., Sitenko,A.G.,
Stepanov,K.N.:1975, Plasma Electrodynamics, Pergamon Press, Oxford.

\end{document}